\documentstyle[graphicx]{mn}

\newif\ifAMStwofonts
\AMStwofontstrue

\def\pg{{PG1211+143}}
\def\ngc{{NGC4051}}

\def\xmm{{\it XMM-Newton}}
\def\chandra{{\it Chandra}}
\def\suzaku{{\it Suzaku}}

\def\et{{et al.\ }}

\newcommand{\ls}{\mathrel{\hbox{\rlap{\hbox{\lower4pt\hbox{$\sim$}}}\hbox{$<$}}}}
\newcommand{\gs}{\mathrel{\hbox{\rlap{\hbox{\lower4pt\hbox{$\sim$}}}\hbox{$>$}}}}

\def\Msun{\hbox{$\rm ~M_{\odot}$}}

\def\rchi{{$\chi^{2}_{\nu}$}}

\def\H0{{\rm ~km~s^{-1}~Mpc^{-1}}}

\def\et{{et al.}}

\title[High velocity outflow of \pg]
        {The soft X-ray spectrum of the luminous narrow line Seyfert galaxy \pg\  - evidence for a second high velocity outflow component}
\author[K.A.Pounds \et]
        {K.A.Pounds\\
Department of Physics and Astronomy, University of Leicester,
Leicester, LE1 7RH, UK\\}

\date{Accepted ; Submitted }
\pagerange{\pageref{firstpage}--\pageref{lastpage}}
\pubyear{2005}
\begin{document}
\maketitle
\label{firstpage}

\begin{abstract}
An \xmm\ observation of the luminous Seyfert galaxy \pg\ in 2001 revealed the first clear evidence for a highly ionised high speed wind (in a non-BAL AGN), with a
velocity of v$\sim$0.09c based on the identification of blue-shifted absorption lines in both EPIC and RGS spectra. A subsequent analysis of EPIC spectra, including
additional absorption lines, led to an upward revision of the wind speed to $\sim$0.14c, while broad band modelling indicated the need for a second, partial covering
absorber to account for continuum curvature and spectral variability.  We show here, in a new analysis of the \xmm\ RGS data, that this additional absorber is detected in
the soft X-ray spectra, with the higher spectral resolution providing a much improved velocity constraint, with v$\sim$0.07c.  Similar variability of the $\sim$0.07c and
$\sim$0.14c outflow components suggest they are physically linked, and we speculate that occurs by the fast (primary) wind impacting on small clumps of higher density, slow
moving matter close to the disc. We show that strong, velocity broadened  soft X-ray emission features, located at the redshift of \pg\, indicate the extended scale of the
ionised outflow.

\end{abstract}


\section{Introduction}

X-ray spectra from an \xmm\ observation of the narrow-line Seyfert galaxy \pg\ in 2001 provided the first detection in a non-BAL AGN of blue-shifted resonance line
absorption corresponding to a sub-relativistic velocity of $\sim$0.09c (Pounds \et\ 2003). Although a much lower velocity was claimed from a separate analysis, principally
based on the RGS soft X-ray data (Kaspi and Behar 2006), the high velocity was confirmed - and adjusted to 0.14$\pm$0.01c - in a re-analysis where the higher energy
resolution of the MOS cameras provided revised line identification (Pounds and Page 2006). Repeated observations with \xmm, \chandra\ and \suzaku\ have since shown the
high velocity outflow to be persistent, but of variable strength (eg Reeves \et\ 2008). Confirmation that the outflow in \pg\ was both massive and energetic - with
potential importance for galaxy feedback - was obtained from the detection of PCygni and other broad emission features in stacking of the 2001, 2004 and 2007 \xmm\ EPIC
spectra (Pounds and Reeves 2007, 2009). 

More recently the examination of archival data from \xmm\ has shown high velocity ionised winds (UFOs) to be relatively common in nearby, bright type 1 AGN (Tombesi 2010, 2011,
2012), a finding supported by a similar analysis of the \suzaku\ archive (Gofford 2013). The frequency of these detections confirms a typically large covering factor and
hence significant mass and kinetic energy in such winds. Indeed, their integrated mechanical energy may be an order of magnitude greater than required to disrupt the bulge
gas in the host galaxy, suggesting much of the energy in a persistent wind must be lost before reaching the star forming region. 

The first evidence of a fast AGN wind shocking with the ISM or previous ejecta, at a radial distance where strong Compton cooling will cause much of the flow energy  to be
lost, has recently been found in an extended \xmm\ study of the narrow line Seyfert 1 \ngc\ (Pounds and Vaughan 2011, Pounds and King 2013). That outcome may be expected
for many of the low redshift AGN in the Tombesi \et\ sample, with the resulting momentum-driven flow being the effective agent of galaxy feedback (King 2003, 2005). 

The evidence for a cooling post-shock flow in \ngc\ was provided by the higher resolution RGS grating spectra, and it is relevant to note that all existing X-ray
detections of high velocity outflows (v $\geq$ 10000 km s$^{-1}$; 0.03c) have been based on blue-shifted absorption lines identified with highly ionised Fe in lower
resolution CCD spectra. In the present paper we re-examine the RGS soft X-ray spectra of \pg\ to seek further definition of this prototype energetic outflow.  

We assume a redshift for \pg\ of $z=0.0809$ (Marziani \et\ 1996). Spectral fitting is based on the {\tt XSPEC} package (Arnaud 1996) and includes absorption  due to the
line-of-sight Galactic column of $N_{H}=2.85\times10^{20}\rm{cm}^{-2}$  (Murphy \et\ 1996). Estimates for the black hole mass in \pg\ range from $3\times 10^{7}$\Msun\
(Kaspi \et\ 2000) to $1.5\times 10^{8}$\Msun\ (Bentz \et\ 2009), with the lower value making the luminosity close to Eddington. It is interesting to note that historical
X-ray fluxes of \pg\ are generally higher than in the \xmm\ era (Bachev \et\ 2009) further strengthening the case for continuum driving of the fast wind in this source
(King and Pounds 2003).

\section{Absorption velocity profile in the soft X-ray spectrum of \pg}

\pg\ was observed by \xmm\ on 2001 June 15, 2004 June 21 and 2007 December 21 and 23.  Here we concentrate on data from the Reflection Grating Spectrometer  (RGS, Den
Herder \et 2001), with effective exposures of $\sim$54 ks, $\sim$47 ks, $\sim$49 ks and $\sim$38 ks, respectively. 

Although individual soft X-ray spectra are rather noisy from these relatively short exposures, several absorption lines were claimed in the original analysis of the 2001
observation (Pounds \et\ 2003), identification with H- and He-like ions of C, N, O and Ne indicating an outflow velocity in the range $\sim$22000-24000 km s$^{-1}$.
However, the individual absorption lines are weak and - as noted in the Introduction - ion by ion modelling by Kaspi and Behar (2006) failed to confirm this high
velocity in the RGS data. In the present analysis we use the velocity transformation method described in Pounds and Vaughan (2011a) which allows stacking of spectra for
multiple resonance lines in order to enhance the visibility of absorption features.

Figure 1 (top panel) shows the composite velocity profile for the 5 principal resonance absorption lines of OVIII, OVII, NVII, NVI and CVI in the 2001 RGS spectra. 
The bin width of
400 km s$^{-1}$ matches the mid-band RGS spectral resolution. The blue-shifted soft X-ray absorption  reported in Pounds \et\ (2003) is clearly detected, with a broad
absorption trough near -22000 km s$^{-1}$ contributing to a poor fit to a level continuum ($\chi^{2}$ of 123 for 85 degrees of freedom).  Addition of a negative Gaussian to
model that absorption (figure 1, mid panel) finds an outflow velocity of 22230$\pm$240 km s$^{-1}$ and width $\sigma$=1020$\pm$250 km s$^{-1}$, with a revised \rchi\ of
94/82. The absorption feature is highly significant, with an f-test random probability of $6\times 10^{-5}$, or $9\times 10^{-4}$ allowing for 35 possible (1$\sigma$
line width) values across the velocity band. A Gaussian search across the velocity plot finds a weaker absorption velocity at 5590$\pm$230 km s$^{-1}$, with width
$\sigma$=570$\pm$240 km s$^{-1}$ (figure 1, mid panel), yielding an overall \rchi\ of 83/79.  The latter absorption is intriguing, being close to the factor 4 reduction
in velocity expected across a strong shock. However, the f-test shows this second feature to be only marginally significant (in the 2001 data) if no prior velocity is
assumed.

The width of the high velocity absorption feature might be due to velocity variations over the data integration period or to physically separate velocity components. We
find compelling evidence for the latter in Section 3, in terms of different ionisation levels, and meanwhile quantify this separation with an additional, narrow
($\sigma$=400 km s$^{-1}$) Gaussian. The second absorption component lies close to the systemic velocity of \pg, indicated in
the lower panel of figure 1 by the symbol G. Allowing for a mid-range uncertainty in the RGS wavelength calibration of $\sim$8 m\AA\ (120 km s$^{-1}$), local absorption
- perhaps in the Galactic halo - appears an interesting possibility. 

A repeat of the above exercise for the 2004 and 2007 RGS data (see Appendix) finds weaker absorption near -22000 km s$^{-1}$, consistent with weaker Fe K
resonance line absorption in the 2004 and 2007 EPIC data (Pounds and Reeves  2007, 2009). An indication of the near-systemic velocity component 
in both 2004 and 2007 profiles strengthens the case for a constant component in the summed RGS data from all 4 \xmm\ observations.

\begin{figure}                                                          
\centering                                                              
\includegraphics[width=6cm, angle=270]{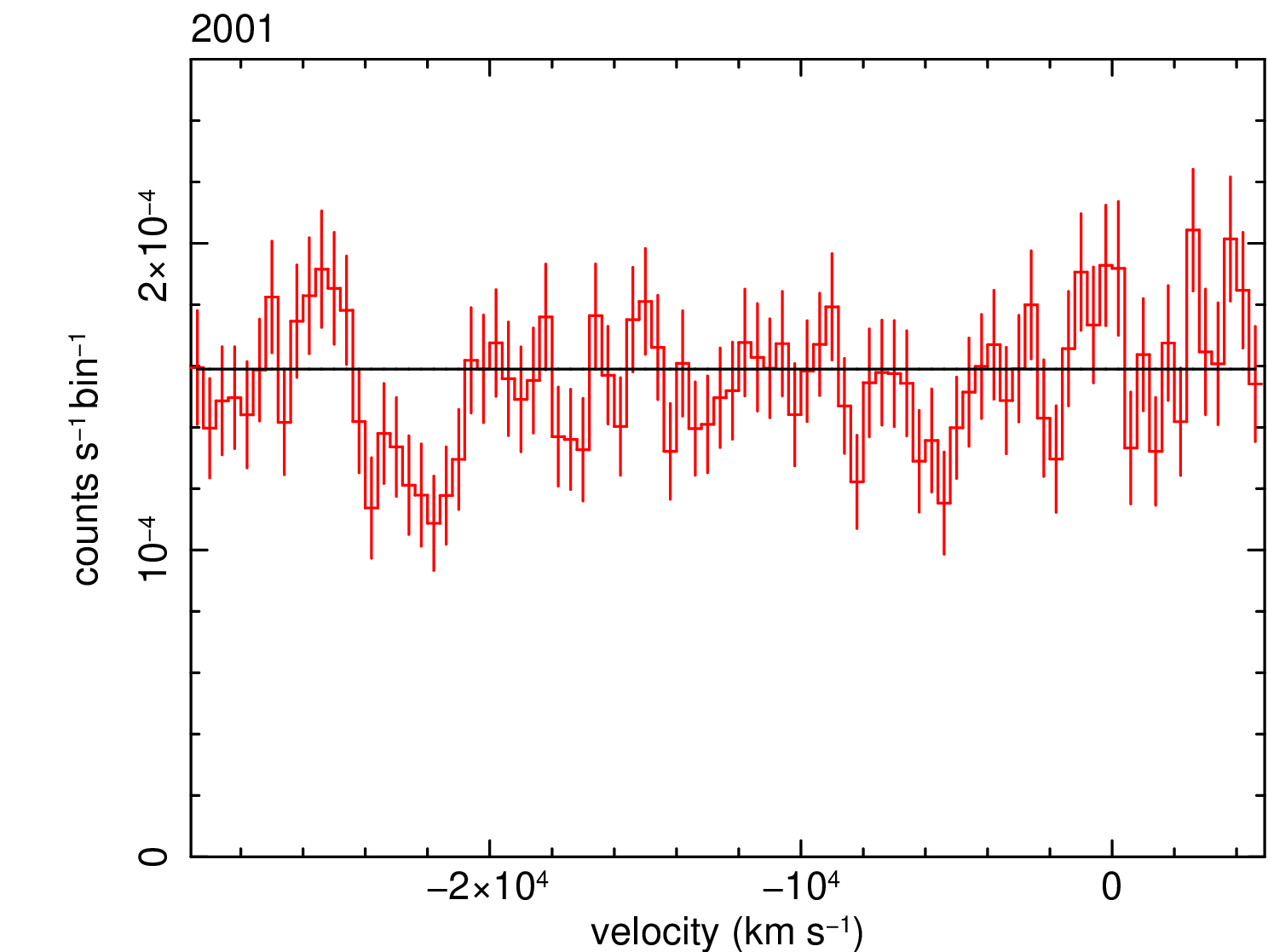}                                          
\centering                                                              
\includegraphics[width=6cm, angle=270]{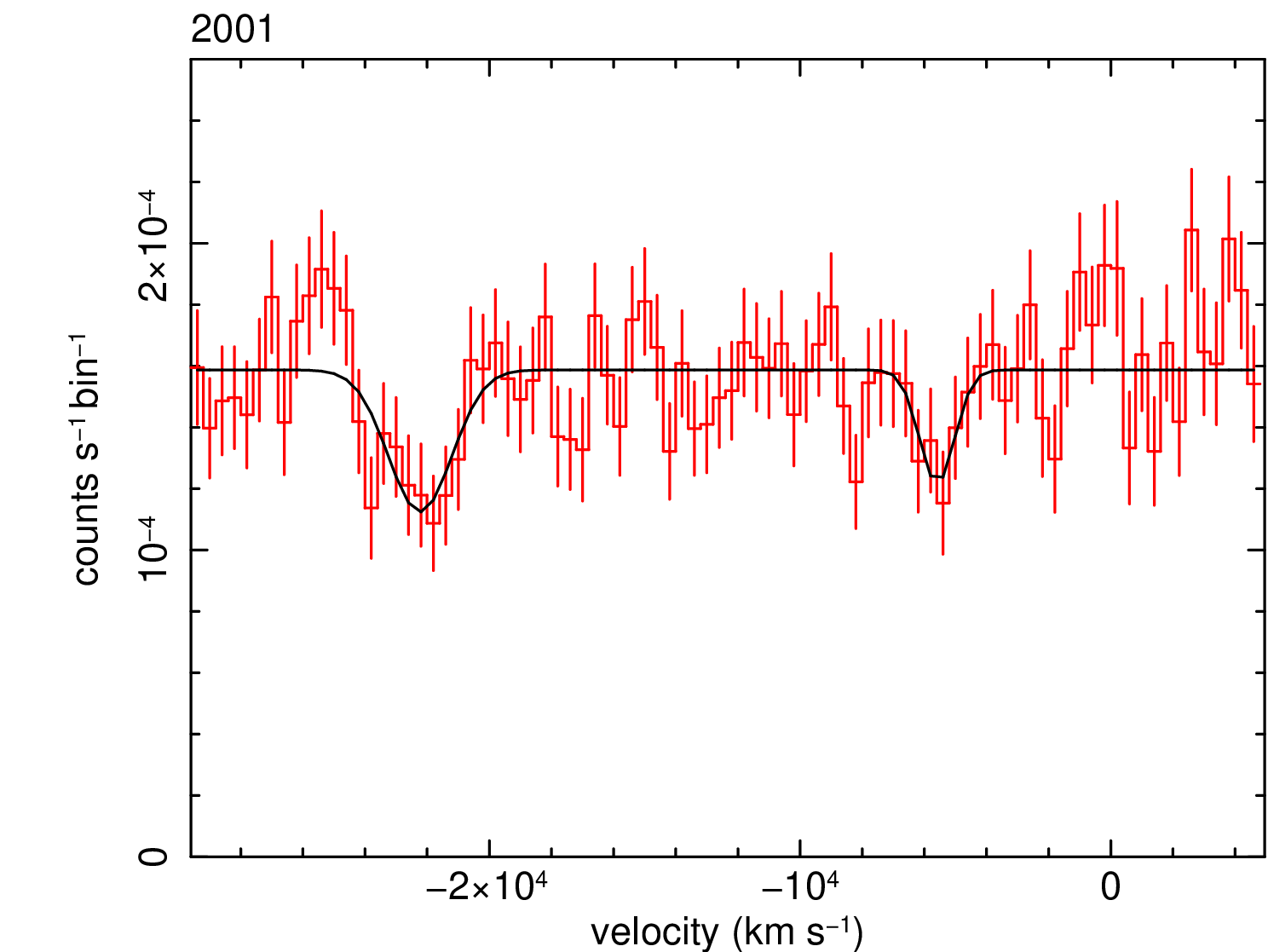} 
\centering                                                              
\includegraphics[width=6cm, angle=270]{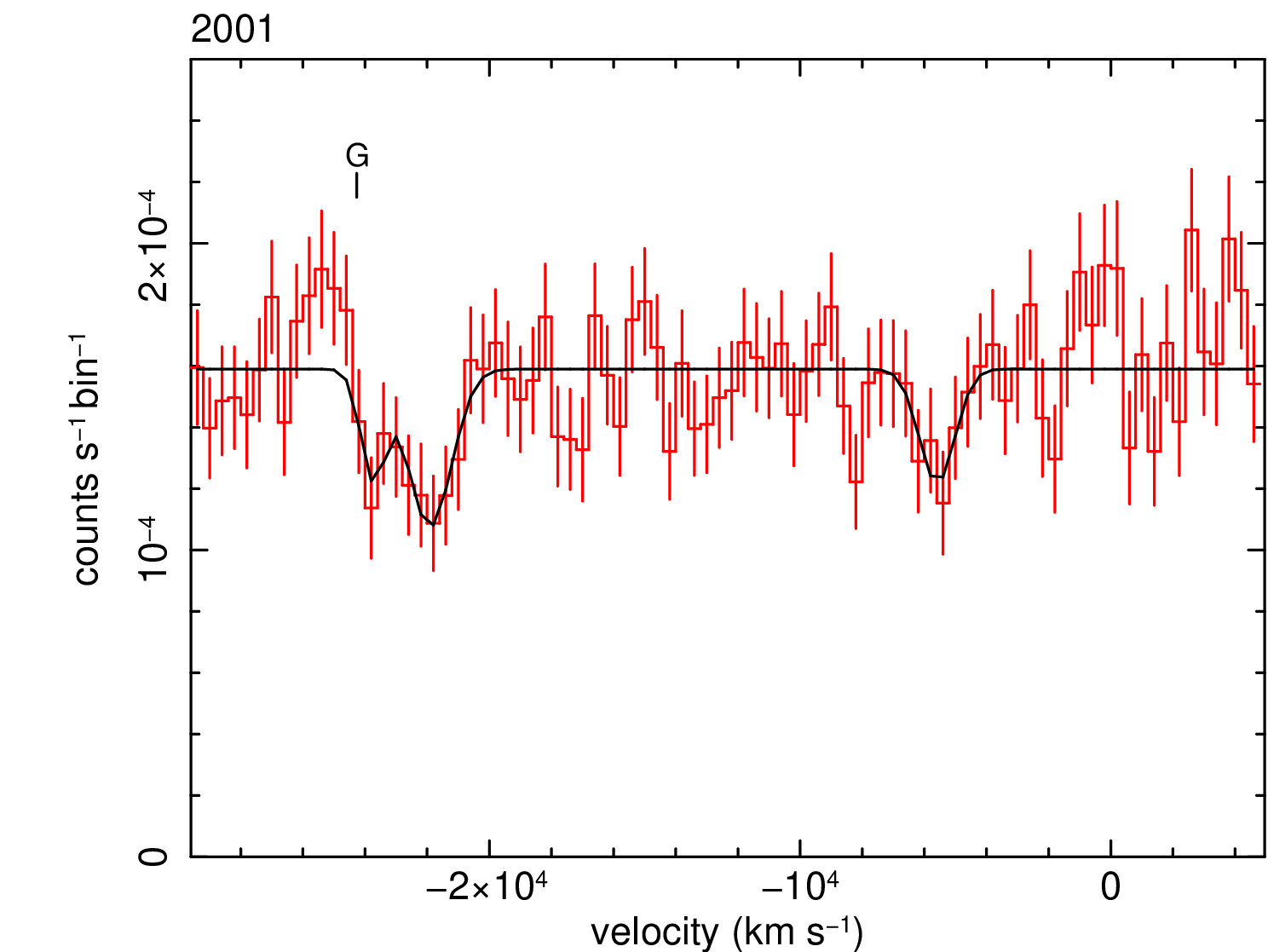}      
\caption                                                                
{(top) Velocity profile for the RGS data from the 2001 \xmm\ observation of \pg, centred at zero velocity in the \pg\ rest frame for the 5 principal resonance lines of
CVI, NVI, NVII, OVII and OVIII. (middle) Gaussian fits to the most significant absorption features correspond to outflow velocities of $\sim$22200 km s$^{-1}$ and
$\sim$5600 km s$^{-1}$.  The higher velocity line is broad with $\sigma$$\sim$1000 km s$^{-1}$. The lower panel resolves the higher velocity component into two
components, one with fixed width of $\sigma$$\sim$400 km s$^{-1}$. Details and statistical significance of the individual Gaussian components are listed in Table 1.
The symbol G indicates the systemic velocity corresponding to the AGN redshift of 0.0809}
\end{figure}  

\begin{table}
\centering
\caption{Parameters of the Gaussian fits to the composite absorption velocity profile in figure 1. Components 1 and 2 refer to the mid panel of figure 1,
with components 1a and 1b to the lower panel fit. Velocities are in km s$^{-1}$ at the AGN redshift}
\begin{tabular}{@{}lccccc@{}}
\hline
Comp & velocity  & width ($\sigma$) & flux ($10^{-5}$ counts & $\Delta$$\chi^{2}$ \\
\hline
1 & -22230$\pm$240 & 1020$\pm$250 & -5.6$\pm$1.6 & 29/3 \\
2 & -5590$\pm$230 & 570$\pm$240 & -3.9$\pm$1.6 & 11/3 \\
1a & -23740$\pm$230 & 400 (f) & -2.9$\pm$1.6 & 6/2 \\
1b & -21910$\pm$190 & 700$\pm$210 & -5.5$\pm$1.6 & 26/3 \\
\hline
\end{tabular}
\end{table}

\section{Comparison with an ionised outflow}

To quantify the properties of the absorption features indicated in the velocity plots (figure 1) the full 2001 RGS spectral data were then modelled with absorption in a
photoionised gas, using grids 18 and 21 from the {\tt XSTAR} code (Kallman et al 1996). Grids 18 and 21 differ in having fixed turbulence velocities of 100 km s$^{-1}$ and
1000 km s$^{-1}$, respectively, with free parameters being column density and ionisation  parameter, with outflow (or inflow) velocities in the AGN rest frame obtained
from the apparent redshift of the absorbing gas. Element abundances were fixed at solar values. In the final fit, described below, grid 21 was preferred for component a,
giving a better constrained column density, with grid 18 being retained for components b and c.

The soft X-ray continuum was first fitted over the full 8-36 \AA\ waveband by a power law plus black body, both attenuated by the Galactic column of
$N_{H}=2.85\times10^{20}\rm{cm}^{-2}$. Positive and negative spectral features in the residuals contributed to a fit statistic \rchi\ of 960/858.  Positive Gaussians were then added to
represent broad line emission of OVII and OVIII and four {\tt REDGE} components (to replicate recombination continua), with parameters determined from the
stacked data in described in Section 4. Re-fitting the 8-36 \AA\ spectrum with adjusted continuum components gave a revised baseline  fit statistic \rchi\ of 920/858.

The addition of 3 photoionised absorbers successively improved the spectral fit to \rchi\ of 854/849. The fit parameters and significance of each {\tt XSTAR}
absorption component are listed in Table 2. For the first two components the fit was started with default values of the variable parameters (log N$_{H}$=21,
log$\xi$=0, z = 0), and repeated trials showed the fitted parameters to be robust. For component c, the initial redshift parameter was set at 0.06 in order to explore the
intermediate velocity indicated in figure 1.   

\begin{figure}                                                          
\centering                                                              
\includegraphics[width=6cm, angle=270]{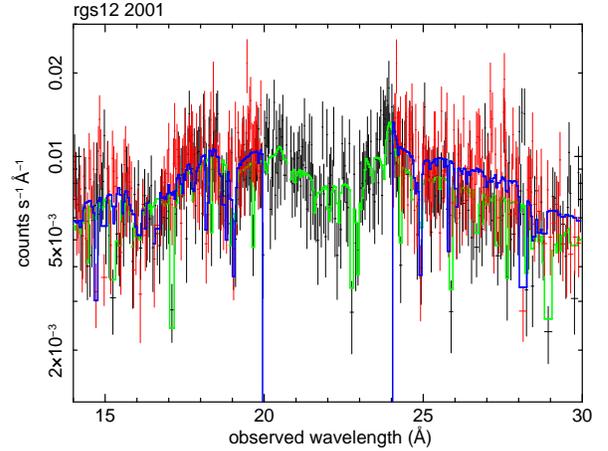}                                           
\caption                                                                
{A section of the RGS spectrum from the 2001 \xmm\ observation of \pg\ with ionised absorption modelled by 3 {\tt XSTAR} components, as listed in Table 2}      
\end{figure}

\begin{table}
\centering
\caption{Parameters of the photoionised outflow fitted to the 2001 RGS data. Column density N$_{H}$ is in H atoms cm$^{-2}$ and ionisation parameter $\xi$ in erg cm s$^{-1}$}
\begin{tabular}{@{}lccccc@{}}
\hline
Comp & log$\xi$ & N$_{H}$  & velocity (km s$^{-1}$) & $\Delta$$\chi^{2}$\\
\hline
a & 2.99$\pm$0.02 & 5$\pm$2$\times10^{22}$ & -21540$\pm$200 & 35/3 \\
b & 0.88$\pm$0.11 & 3$\pm$1$\times10^{20}$ & -23170$\pm$300 & 27/3 \\
c & 1.2$\pm$0.8 & 3$\pm$4$\times10^{19}$ & -6140$\pm$550 & 6/3 \\
\hline
\end{tabular}
\end{table}

The two most significant absorption components in the photoionised modelling (a and b) are both of high velocity (relative to the AGN), with values close to the
strongest absorption feature in the velocity plot of figure 1. However, those {\tt XSTAR} absorbers have very different ionisation parameters and column densities,
which  might indicate that the fast outflow has entrained cooler gas, albeit with a substantially lower column density. An interesting alternative, given the proximity
to the systemic velocity of \pg, is that the lower ionisation component (b) is not associated with the AGN, but arises from low redshift matter in the same line of sight.

While the 3rd photoionised absorption component is of lower statistical significance, it is again consistent with a velocity absorption feature in Figure 1. Although
the ionisation parameter of component 3 is typical of a `warm absorber' the velocity would be unusually high. We return to this point in Section 7.

\section{Evidence for velocity broadened emission}

In an attempt to quantify the covering factor/collimation of the fast outflow in \pg\ Pounds and Reeves (2007) modelled the broad band EPIC spectrum with photoionised absorption and emission, assuming
both to arise from the same ionised flow. An interesting factor in achieving a good fit was a requirement to broaden the emission lines  by inclusion of a Gaussian smoothing parameter ({\tt GSMOOTH} in
{\tt XSPEC}). In addition, the {\tt XSTAR} emission spectrum indicated a mean velocity (in the AGN rest frame) of $\sim$3000$\pm$3000 km s$^{-1}$, locating the ionised gas observed in absorption close
to the redshift of the AGN. Here we examine individual emission features obtained by stacking the soft X-ray spectra from all four RGS observations to clarify the above indications. As reported
previously, the only strong emission features are broad. 

Figure 3 (top) shows coarsely binned RGS 1 spectra summed over all four \xmm\ observations, plotted as a ratio to a power law plus black body continuum. Five broad
emission features are modelled with positive Gaussians to determine the approximate wavelength and width of each feature, being found - for all 5 - to lie close to the
respective wavelength (adjusted for the AGN redshift) of OVIII Lyman-$\alpha$, the OVII triplet and the radiative recombination continua (RRC) of OVIII, OVII and CVI. This is an
important confirmation that the ionised gas seen in absorption is indeed associated with the AGN, while the width of the broad OVIII Lyman-$\alpha$
emission line would correspond to velocity broadening of $\sim$20000 km s$^{-1}$ (FWHM), a value similar to that found for the PCygni line observed in Fe K (Pounds
and Reeves 2009), and consistent with a wide angle outflow.

To better quantify the individual emission features, with the RRC widths potentially containing important information on the related electron temperature, the stacked
\xmm\ soft X-ray data were then modelled in {\tt XSPEC}, where each RRC (including NeIX) was more appropriately fitted with a {\tt REDGE} component.  

In the lower panel of figure 3 the {\tt XSPEC} fit is reproduced, including four significant RRC and broad and narrow emission components of OVIII Lyman-$\alpha$ and the
OVII triplet.  The component parameters of the main broad emission features and corresponding statistical improvement over the baseline continuum fit are listed in Table
3. 

\begin{figure}                                                          
\centering                                                              
\includegraphics[width=6cm, angle=270]{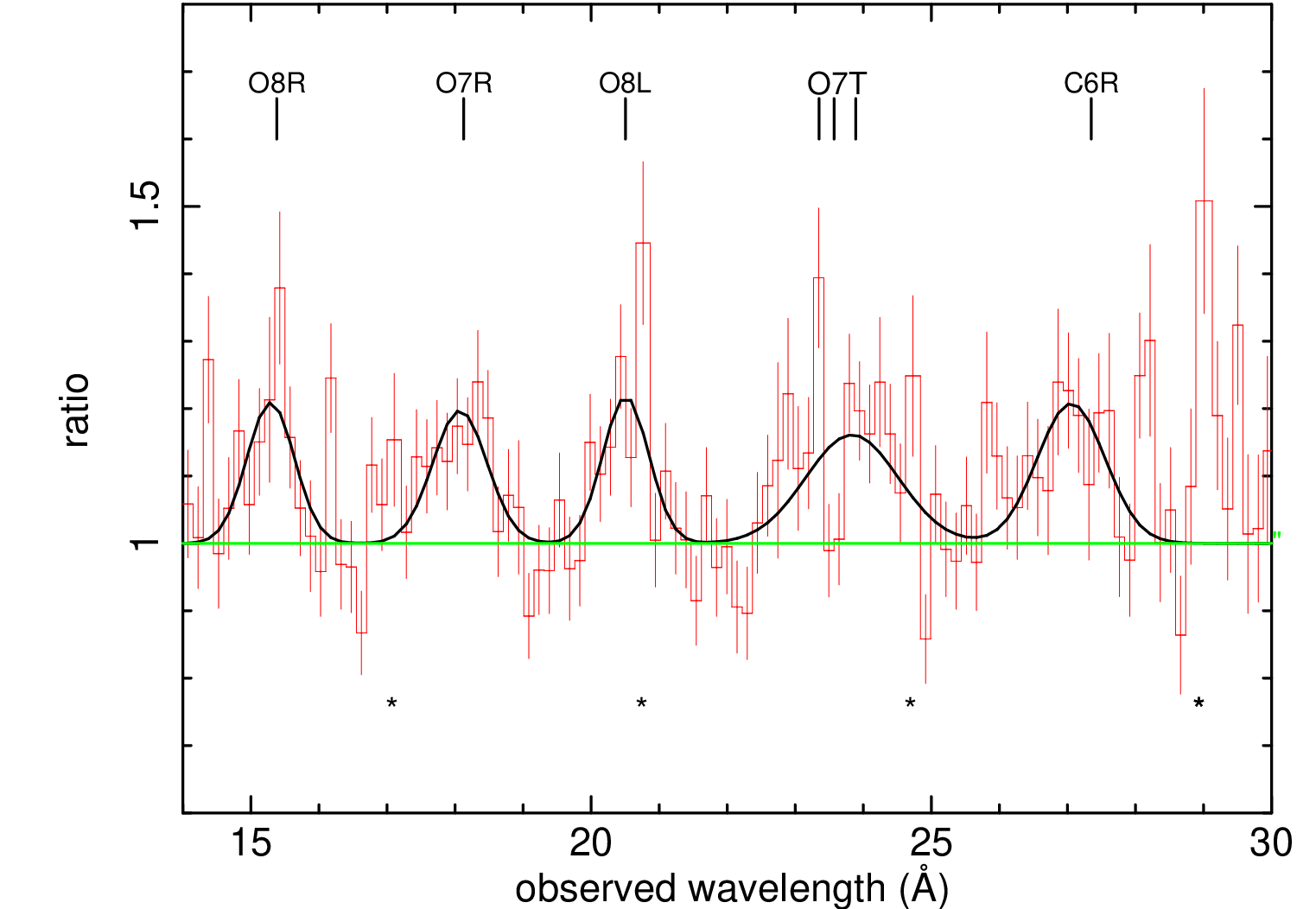}                                           
\centering                                                              
\includegraphics[width=6cm, angle=270]{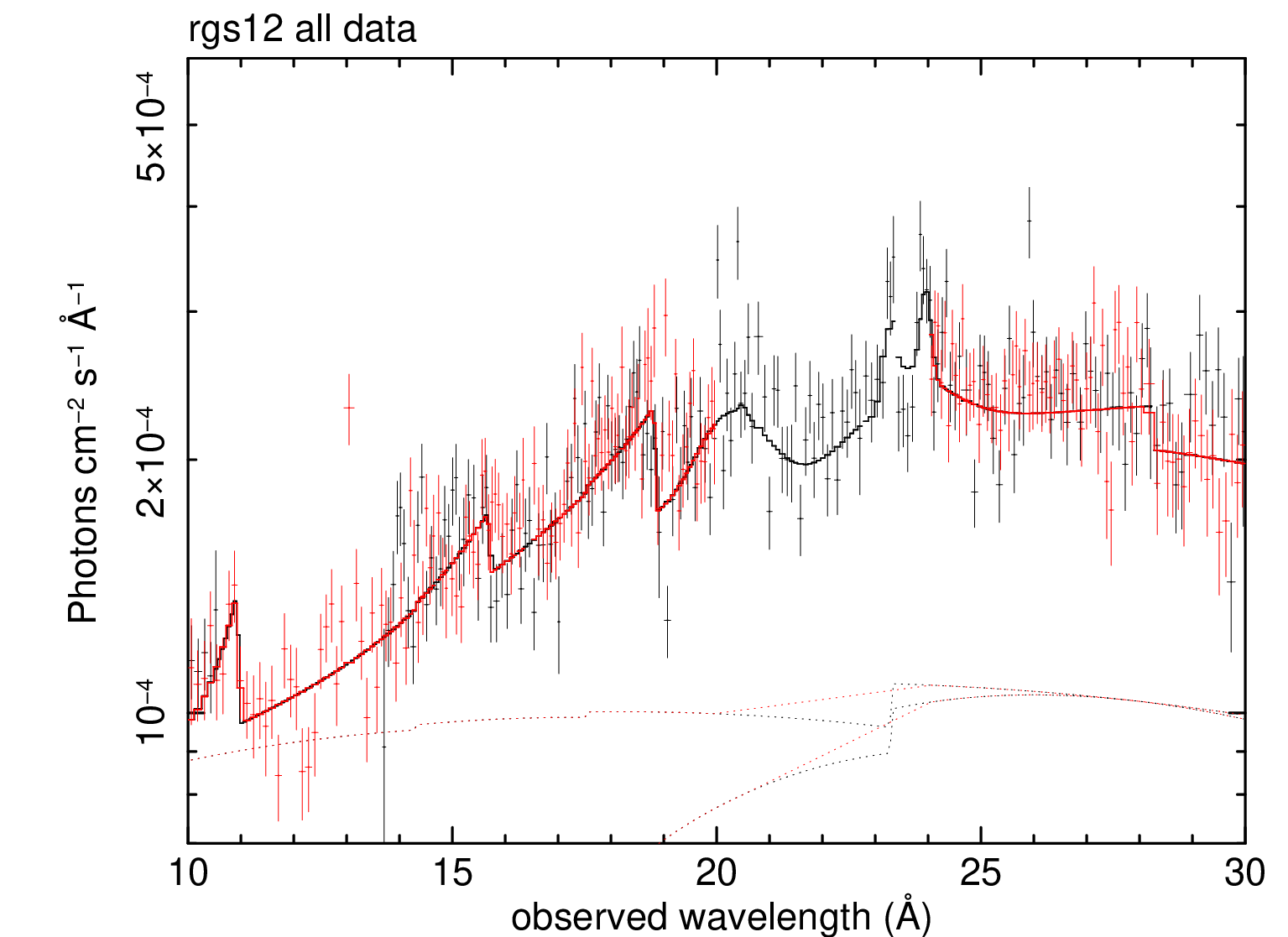} 
\caption                                                                
{(top) Coarsely binned RGS 1 data summed over all four \pg\ observations plotted as a ratio to a power law plus black body continuum
fit. Five broad emission features are modelled with positive Gaussians, with respective wavelengths and widths listed in Table 3. The rest wavelengths of OVIII
Lyman-$\alpha$, the  OVII triplet, and the RRC of NeIX, OVIII, OVII, and CVI are included for comparison. (lower) The same composite spectrum including emission line 
and RRC components with parameters from {\tt XSPEC fitting}}      
\end{figure}  

\begin{table}
\centering
\caption{Parameters of the OVIII Lyman-$\alpha$ broad emission line and RRC of NeIX, OVIII, OVII and CVI fitted to the combined RGS data from \xmm\ observations in
2001, 2004 and 2007. Component fluxes are in $10^{-5}$ photons cm$^{-2}$ s$^{-1}$. Observed and laboratory line and RRC threshold energies are listed in the rest
frame of PG1211+143}
\begin{tabular}{@{}lcccccc@{}}
\hline
Comp & obs(keV) & lab(keV) & $\sigma$/kT(eV)& flux & $\Delta$$\chi^{2}$\\
\hline
O8 L-$\alpha$ & 0.66$\pm$0.01& 0.654 & 19$\pm$3 & 9$\pm$2 & 7/3\\
Ne9 RRC & 1.22$\pm$0.01 & 1.196 & 66$\pm$28 & 2.2$\pm$0.6 & 11/3\\
O8 RRC & 0.85$\pm$0.01 & 0.872 & 58$\pm$16 & 3.0$\pm$0.6 & 21/3\\
O7 RRC & 0.72$\pm$0.01 & 0.739 & 49$\pm$6  & 10$\pm$1 & 76/3\\
C6 RRC & 0.48$\pm$0.01 & 0.490 & 38$\pm$10  & 11$\pm$3 & 16/3\\
\hline
\end{tabular}
\end{table}

Component energies are adjusted for the redshift of \pg\ in the table to allow direct comparison with theoretical values. The agreement is good, with the mean energy
of the broad Lyman-$\alpha$ line again suggesting a small blueshift. Assuming  the RRC are affected by similar velocity broadening to the OVIII Lyman-$\alpha$ line,
yields an electron temperature of the relevant gas component of kT$\sim$10--20 eV,  consistent with the fast outflow being close to the Compton temperature in the
radiation field of the AGN. (The derived value is consistent with an unpublished OVII RRC electron temperature of 14$\pm$5 eV from the \chandra\ LETG observation in 2004;
Reeves private communication).

\section{Scattering from the fast ionised wind}

The emission features described above provide an important constraint on the scale of the outflow at v$\sim$0.07c.  

The {\tt XSPEC} fit to the summed RGS data finds a flux for the strong OVII RRC of $\sim$$10^{-4}$ photons cm$^{-2}$ s$^{-1}$. Assuming a temperature of the  
soft X-ray
emitting gas of kT$\sim$10 eV, the relevant recombination rate is of order $10^{-11}$ cm$^{3}$ s$^{-1}$ (Verner and Feldman 1996). For a solar abundance of oxygen and 
parent ion fraction of 0.3, we find an emission measure of $\sim$$3\times10^{66}$cm$^{-3}$, for a redshift-distance to \pg\ of 350 Mpc. 

We assume the wind extends from launch at the escape velocity radius and coasts at near-constant velocity before colliding with the
ISM. King and Pounds (2013) suggest that will typically be where the initial ISM has been swept up by
radiation pressure into an optically thin shell. For \pg\ the predicted transparency radius $R_{tr}$ $\sim$8 pc, yielding an emission  volume ($4b\times R^{3}$) of
$\sim$$6b\times10^{58}$ cm$^{3}$. For a flow collimation b (=$\Omega/4\pi$) of 0.5 (Pounds and Reeves 2009), comparison with the above emission measure then implies a
mean electron density of $\sim$$10^{4}$ cm$^{-3}$. While that would give a column density over 8pc a factor $\sim$5 larger than component 1 in table 2, the latter is a
crude estimate for a coasting wind where the particle density will fall as $r^{2}$ and the absorbing column is dominated by the inner radii. 

The strength of the broad soft X-ray emission features, and their lack of obvious variability over several years, supports the view the corresponding ionised outflow
continues unchecked to a relatively large distance. 

\section{Comparison with previous EPIC analyses}

As noted in the Introduction, previous \xmm\ studies of the outflow in \pg\ have been primarily based on the EPIC spectra. Although exhibiting a strong `soft excess'
over a `primary' power law continuum ($\Gamma$$\sim$2.2), RGS spectra show only weak narrow spectral features.  Detailed modelling of pn and MOS spectra (Pounds and
Reeves 2007) suggested an explanation, with velocity-broadened emission lines and soft X-ray absorption lines being `diluted' by a second, softer and
less absorbed, continuum component ($\Gamma$$\sim$3.2).

Broad-band spectral modelling required two ionised absorbers in the partial covering fit, with log$\xi$$\sim$1.5 and $\sim$2.9. Outflow velocities  from the {\tt XSTAR}
modelling were, respectively, $\sim$0.07$\pm$0.02c and $\sim$0.14$\pm$0.01c, with the higher velocity, higher ionisation absorption evidently driven by fitting blue-shifted resonance
lines of Fe K and other heavy metal ions (as in Pounds and Page 2006), while the lower ionisation, lower velocity absorption was primarily responsible for the
continuum curvature at lower energies, with the velocity determination consequently less secure. 

Finding an outflow velocity of $\sim$0.07c in the 2001 RGS data now greatly strengthens the case for two high velocities in the \pg\ outflow, with a physical association being indicated 
by Fe K resonance lines and broad band
absorption also being more pronounced in 2001 (eg Figure 2 in Pounds and Reeves 2009). The absence of absorption in the RGS data  at the higher 
velocity of $\sim$0.14c can be explained by that component being more highly ionised. 

We discuss the implications that both $\sim$0.07c and $\sim$0.14c flow components were present during the 2001 observation in Section 8.

\section{Absorption in matter at lower redshift}

Component b in the Xstar modelling of the 2001 RGS data, evidently linked with the Gaussian component 1a in Table 1, provides the first evidence of absorption in matter
perhaps not associated with \pg. The persistence of a narrow absorption feature close to the systemic velocity in the 2004 and 2007 RGS data (figure 4) lends support to
a local origin. However, the derived velocity (Tables 2 and 3) is $\sim$500-1000 km s$^{-1}$ (2 and 3 $\sigma$) less than the systemic velocity, arguing against an origin in
the Galaxy, although Herenz \et (2013) report significant CIV absorption in high velocity Galactic Halo clouds receding  at 169 and 184  km
s$^{-1}$ in the line of sight to \pg. 

The reality of component c in the {\tt XSTAR} modelling is also backed by a similar velocity (relative to the AGN) of component 3 in figure 1, and might be the
first evidence for a warm absorber in \pg. The factor 4 difference in velocity compared to component `a' would be consistent with the recent suggestion
that the warm absorber arises from gas accumulating after the fast wind impacts the ISM or previous ejecta (Pounds and King 2013). 

An interesting alternative explanation for component c is absorption in dwarf galaxies along the line of sight to \pg. Prochaska \et\ (2011) report OVI absorbing column
densities of $\sim$2$\times 10^{14}$ cm$^{-2}$ z$\sim$0.0510 and z$\sim$0.0645. Component 3 in table 2 would give an OVI column only a factor of a few greater, while a
small range in ionisation parameter would provide consistency.     

\section{Discussion}

Establishing a second high velocity outflow component in \pg\ is the most important result of the present analysis, being the first example, to our knowledge, of {\it simultaneous} 
multiple high velocities in a UFO.
Although it is relatively common for broadband X-ray spectra of AGN to be fitted with a partial covering model, no velocity information has been 
reported for the partial covering absorber. 

Detection of multiple high velocities will, of course, affect the overall energy and momentum budget, with important relevance to feedback. We restrict discusion here to
considering how these velocity components may be physically linked. 

Conceivably, multiple speeds could arise from different radii of the inner disc, with corresponding escape velocities, or reflect real variability over the duration of an observation. 
However, for continuum driving that would require (King 2010, equ. 8) a factor two change in accretion rate over a few hours in the 2001 observation, which may seem unlikely.  

An interesting alternative, given the importance of obtaining a better understanding of the fate of UFOs as they move out into the surrounding medium, is  that the lower velocity,
lower ionisation component represents the collision of two parcels of gas with similar masses, where one has the primary wind speed and the other is more or less static until the
collision. Once a similar wind mass has shocked, the speed is  automatically $\sim$ v/2. An important constraint on this idea is that `similar mass' requires a `similar column density',
suggesting the
collision must occur at a relatively small radius for a radially expanding wind.

\section{Summary}

A re-analysis of soft X-ray spectra of the luminouus Seyfert galaxy \pg\ has confirmed that the high velocity outflow observed in highly ionised Fe K absorption lines is also
evident in soft X-ray lines. In particular, the absorption  velocity profile obtained by combining spectra near multiple resonance lines finds apparent outflow velocities (wrt
the AGN) of $\sim$23730$\pm$230 and 21910$\pm$190 km s$^{-1}$.   

Spectral fitting with the {\tt XSTAR} photoionisation code (Kallman \et 1986) confirms the existence of two separate absorption components, with similar apparent velocities but
very different ionisation levels. The similarity of the low ionisation soft X-ray component velocity to the systemic velocity of \pg\ (24270 km s$^{-1}$) raises the interesting
possibility of absorption in local
matter in line of sight to the AGN, with that local interpretation being supported by the narrow absorption feature persisting in the 2004 and 2007 data. 

In contrast, the $\sim$22000 km s$^{-1}$ absorption is clearly linked to the high velocity wind in \pg, comparison with previous analyses of EPIC data finding similar
variability in both Fe K resonance line and continuum absorption, both being strongest in 2001. 
Intriguingly, broad band spectral modelling in Pounds and Reeves (2007, 2009) found outflow velocities of $\sim$0.07c and $\sim$0.14c, with the lower value being driven by the
continuum curvature below $\sim$2 keV. 

While the present soft X-ray analysis provides a more secure measure of the $\sim$0.07c flow component, the higher velocity is not
detected, presumably due to that flow component being too highly ionised for significant soft X-ray opacity. Establishing a direct physical link between the two high velocity absorbers will require
more extended observations, but the velocity ratio does offer the interesting possibility that the lower velocity matter is being entrained when the fast wind impacts on the ISM
or slow moving ejecta. While the impact will shock, with likely loss of mechanical energy by Compton cooling, the momentum of the residual flow will be conserved, with a factor
2 velocity difference  perhaps most readily detectable.

Although the presence of discrete `clumps' of low ionisation matter has been proposed to explain rapid, spectrally neutral flux changes (eg. Brennenman \et\ 2013), the present
analysis would provide the first confirmation of a high radial velocity for low ionisation matter, with implications for both the wind acceleration and outflow energetics.  

Coarse binning of the summed soft X-ray spectra together with {\tt XSPEC} spectral fitting confirms the velocity-broadened soft X-ray emission features indicated in the earlier EPIC
analyses, with resonance line and RRC fluxes being interpreted as scattering of the AGN X-ray continuum from a wide angle fast outflow, with an emission measure consistent with
the wind extending  to a relatively large radius. 

Finally, a weak absorption feature at $\sim$5600-6100 km s$^{-1}$ may represent the first detection of a Warm Absorber in \pg.  The weakness of such soft X-ray features and the
absence of  UV absorption (Ganguly \et\ 2013) may suggest the outflow in \pg\ remains relatively highly ionised even after the anticipated shocking with the ISM or previous ejecta 
(Zubovas and King 2012).

Substantially deeper soft X-ray observations of ultra fast outflows, such as that in \pg, will have considerable potential in clarifying 
the ionisation and dynamical structure in powerful ionised winds in AGN, and their subsequent interaction with the ISM or slower moving ejecta.

\section*{ Acknowledgements }
The author is pleased to acknowledge a continuing and stimulating dialogue with Andrew King attempting to clarify and understand the properties of AGN winds. The results reported here are based on observations obtained with \xmm, an ESA science mission with instruments and contributions
directly funded by ESA Member States and the USA (NASA).

\section{Appendix}

In Section 2 the composite velocity profile for the 5 principal resonance absorption of CVI, NVI, NVII, OVII and OVIII from the 2001 RGS data showed a
significant broad high velocity feature which may be a blend of blue-shifted absorption at $\sim$22200 and $\sim$23700 km s$^{-1}$, the latter being close to the systemic
velocity for the redshift of \pg. The existence of 2 separate absorption components was supported by the XSTAR modelling in Section 3, which finds a lower ionisation
parameter for the near systemic velocity. 

Figure 4 compares the same composite velocity profiles for the sum of the 2004 and 2007 RGS data where the the broad absorption near -22000 km s$^{-1}$ is much less evident,
while the narrow component close to the systemic velocity remains, as does the absorption feature near -5600 km s$^{-1}$ (in the AGN rest frame). The lower panel of figure
4 shows the same profile for the combined 2001, 2004 and 2007 RGS data, again binned at 400 km s$^{-1}$. 

While it is difficult to assess such weak features, evidence for the -22000 km s$^{-1}$ feature arising from high velocity soft X-ray absorption in the highly ionised wind
of \pg\ is strengthened by the its variability, being much weaker in the later observations (as were the continuum and resonance line absorption in the EPIC spectra).
The consistent velocities and strengths of both the near-systemic absorption feature and that -5600 km s$^{-1}$ supports their reality. While the latter
feature might be the first evidence for a warm absorber in \pg\, the velocity would be unusually high (Tombesi \et\ 2013), while an intriguing alternative would be ionised gas
associated with a pair of dwarf
galaxies showing strong OVI absorption at cz= 0.0510 and 0.0645 Prochaska \et\ (2011).

\begin{figure}                                                          
\centering                                                               
\includegraphics[width=6cm,angle=270]{2001b.ps}                                           
\centering                                                              
\includegraphics[width=6cm, angle=270]{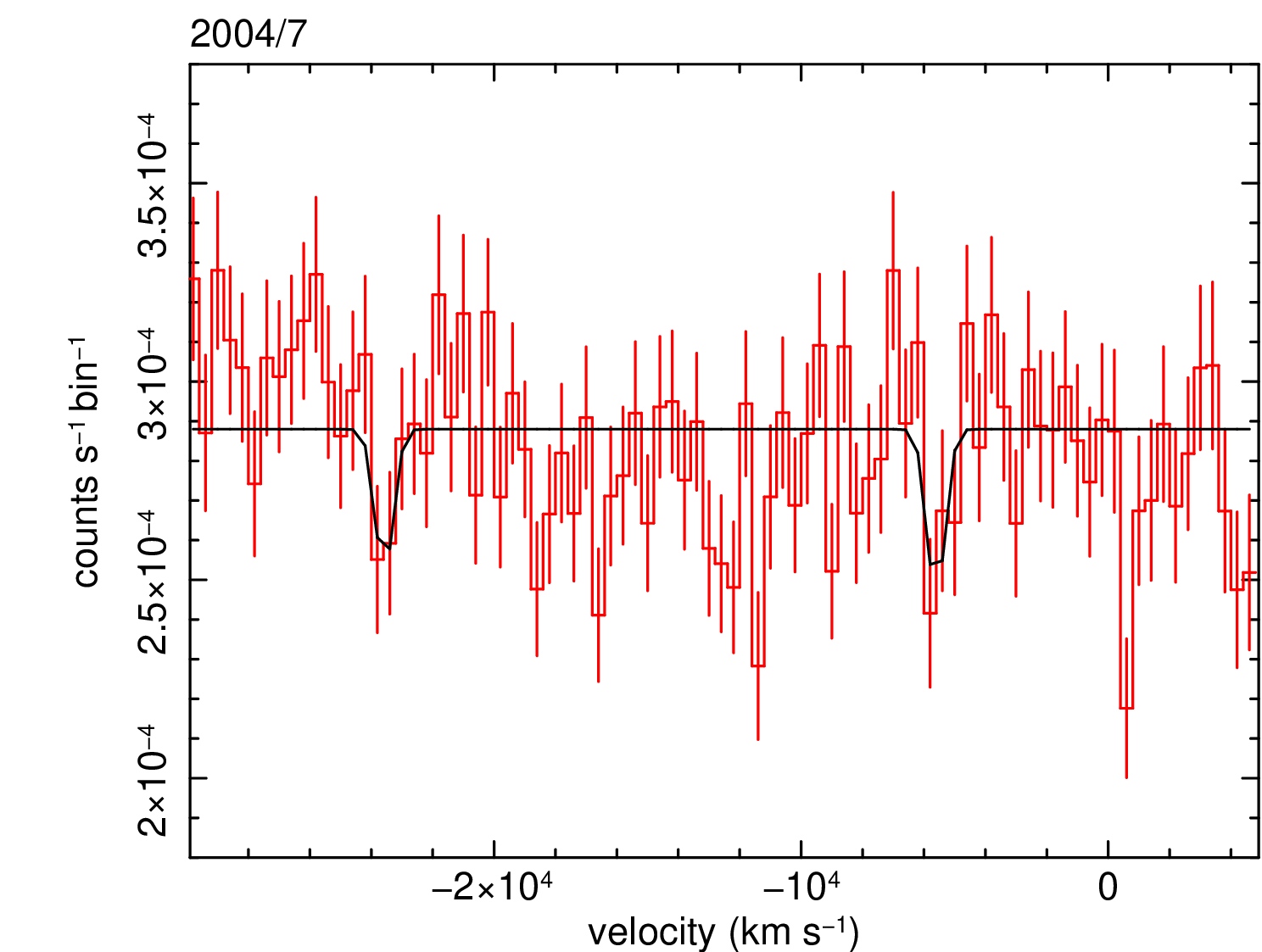}  
\centering                                                              
\includegraphics[width=6cm, angle=270]{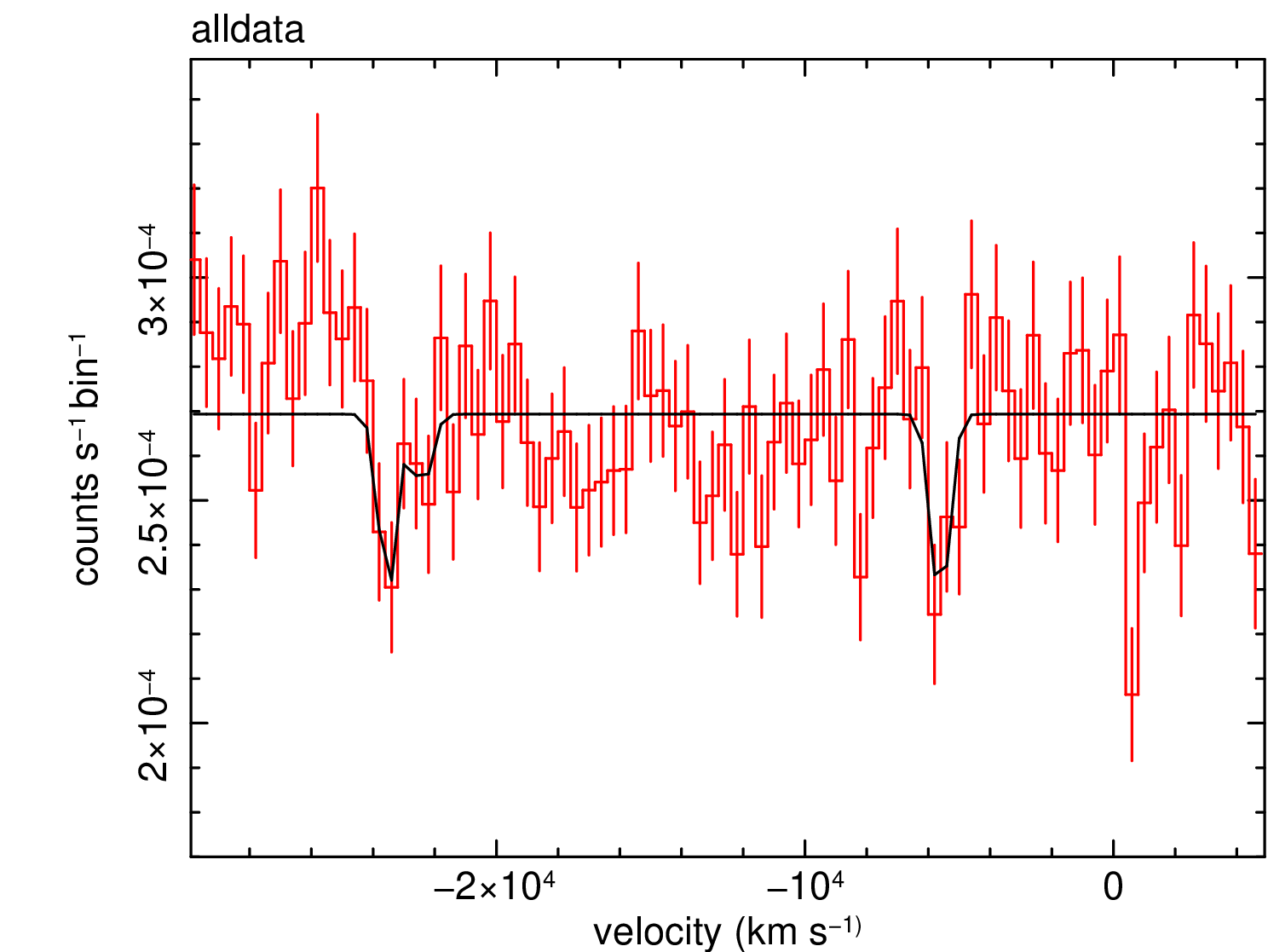}       
\caption                                                                
{(Comparison of the 5-line velocity profile for the 2001 RGS observation of \pg\ with that for the summed 2004 and 2007 data and for the sum of all four \xmm\
observations.  The strong absorption at $\sim$22200 km s$^{-1}$ in 2001 is much weaker in the later observations (as were the continuum and resonance line absorption in the
EPIC spectra), while narrow  absorption features close to the systemic velocity of \pg\ and to absorption from a line-of-sight dwarf galaxy are enhanced in the all\_data profile.
Note the change  of scale in the lower plots}  
\end{figure}  
  
\begin{table}
\centering
\caption{Parameters of the Gaussian fits to the composite absorption velocity profile in figure 4 (lower panel). Component numbers are as in Table 1. Velocities are in 
km s$^{-1}$ and for the redshift of \pg}
\begin{tabular}{@{}lcccc@{}}
\hline
Comp & velocity  & width ($\sigma$) & flux ($10^{-5}$counts)& $\Delta$$\chi^{2}$ \\
\hline
1a & -23740$\pm$270 & 400(f)& -4$\pm$1.3 & 7/2 \\
1b & -22400$\pm$420 & 400(f) & -1.7$\pm$1.3 & 3/2 \\
2 & -5580$\pm$150 & 400(f)  & -4.4$\pm$1.4 & 7/2 \\
\hline
\end{tabular}
\end{table}

\end{document}